\documentclass[a4paper,11pt]{article}
\usepackage{fix-cm}
\usepackage{jinstpub} 
\usepackage{lineno}
\usepackage{eurosym}
\usepackage{siunitx}
\usepackage{array}
\usepackage{amsmath,amsfonts}
\usepackage[caption=false,font=normalsize,labelfont=sf,textfont=sf]{subfig}
\usepackage{textcomp}
\usepackage{orcidlink}

\title{Cryogenic cesium iodide as a potential PET material}

\author[a,b,1]{S. R. Soleti\orcidlink{0000-0002-5526-1414},\note{Corresponding author.}}
\author[a]{A. Castillo,}
\author[a,c]{J. Collar\orcidlink{0000-0002-0650-0626},}
\author[a,d]{M. del Barrio-Torregrosa,}
\author[a]{C.~Echeverria,}
\author[a,d]{M. Seemann\orcidlink{0009-0006-9813-7305},}
\author[a]{D. Zerzion,}
\author[a,b]{J.  J. G\'omez Cadenas\orcidlink{0000-0002-8224-7714}}
\affiliation[a]{Donostia International Physics Center, \\San Sebasti\'an / Donostia, E-20018, Spain}
\affiliation[b]{Ikerbasque (Basque Foundation for Science), \\Bilbao, E-48009, Spain}
\affiliation[c]{Enrico Fermi Institute, Kavli Institute for Cosmological Physics, and Department of Physics\\
University of Chicago, \\Chicago, Illinois 60637, USA}
\affiliation[d]{Universidad del Pais Vasco (UPV/EHU), \\Bilbao, E-48009, Spain}

\emailAdd{roberto.soleti@dipc.org}

\abstract{Total Body PET (TBPET) scanners have recently demonstrated the ability to significantly reduce both acquisition time and the administered radioactive dose, thanks to their increased sensitivity. However, their widespread adoption is limited by the high costs associated with the current available systems. 

Cesium iodide (CsI), though historically less favored for PET due to its lower stopping power and light yield compared to crystals like LYSO, shows remarkable improvement when operated at cryogenic temperatures ($\sim$100 K). Under these conditions, CsI light yield rises dramatically to about 100 photons/keV, providing excellent energy resolution and good coincidence time resolution at a lower cost — typically 3 to 5 times cheaper than other crystals at parity of radiation length.

In our study, we measured the light yield, the energy resolution and the coincidence time resolution as a function of temperature for two pure CsI crystals read out by a pair of silicon photomultipliers (SiPMs). An energy resolution of 6.3\% FWHM and a coincidence time resolution of 1.84~ns at 511 keV were achieved at a temperature of 104~K. These results point towards the potential of cryogenic CsI as a cost-effective, high-performance material for TBPET scanners.}


\arxivnumber{2406.13598} 

\begin{document}
\maketitle
\flushbottom

\section{Introduction}
Positron emission tomography (PET) is a powerful imaging technique that plays an important role in medical diagnostics, employing positron-emitting radionuclides to track biologically active molecules. The annihilation of positrons and electrons produces gamma rays of 511~keV, which are detected to create detailed images of functional processes within the body.

Most modern commercial PET scanners employ lutetium-based scintillating crystals, such as lutetium yttrium oxyorthosilicate (LYSO). This material has optimal performances for PET imaging, providing an excellent stopping power, a good light yield and a fast decay time. However, due to its high cost (45-50 \euro/cm$^3$), as well as the cost of the associated electronics and readout, commercial scanners are typically limited to small axial lengths and are thus able to cover only a small portion of the patient’s body at a time~\cite{vandenberghe2023potential}. 

Although large TBPET scanners have been built in the last decade, showing an impressive increased performance (e.g., United Imaging uEXPLORER~\cite{cherry2018total}, Siemens Biograph Vision Quadra~\cite{prenosil2022performance}), the cost of such apparatus is beyond the possibilities of most hospitals. 

Therefore, developing a cost-effective TBPET system will require identifying a less expensive alternative to lutetium-based crystals, ideally one that provides similar performance.
In this context, the possibility of using scintillating crystals based on cesium iodide (CsI) is worth being explored. Table~\ref{tab:crystals} shows the properties of CsI compared with other crystals commonly used for PET. At parity of radiation length, CsI is five times less expensive than LYSO and two-three times less expensive than BGO. Additionally, its raw materials are widely available, and crystal growth requires a significantly lower temperature -- approximately 800 $^{\circ}$C, compared to the 1000 $^{\circ}$C and 2000 $^{\circ}$C needed for BGO and LYSO, respectively~\cite{singh2024bright}. 
\begin{table}[htbp]
\centering
\small
\caption{Properties of scintillating materials commonly used for PET, compared with scintillators based on cesium iodide. Data for LYSO, BGO and CsI(Tl) have been obtained from vendor~\cite{Luxium}. Cost is normalized by one radiation length $X_0$.}
\label{tab:crystals}
\smallskip
\begin{tabular}{ c | c | c | c | c | c | c | c }

\hline
Material & $Z_{eff}$ & $X_0$ (cm) & $\rho$ (g/cm$^3$) & LY ($\gamma$/keV) & $\tau$ (ns) & Peak (nm) & Cost (\euro/cm$^3\cdot X_{0}$) \\
\hline
LYSO         & 66 & 1.14 & 7.4 & 30  & 53  & 420 & 50 \\
BGO          & 74 & 1.12 & 7.1 & 10  & 300 & 480 & 28 \\
CsI(Tl)      & 54 & 1.86 & 4.5 & 55  & 1000 & 560 & 10 \\
CsI (300~K) & 54 & 1.86 & 4.5 & 5   & 15  & 310 & 10 \\
CsI (100~K) & 54 & 1.86 & 4.5 & 100 & 800 & 350 & 10 \\

\hline
\end{tabular}
\end{table}

Pure CsI has been adopted in particle physics for experiments set in high-rate and high-radiation environments, given its fast decay time and radiation hardness~\cite{Atanov:2016hoz}. At first glance, its low light yield (5~photons/keV), UV emission, average radiation length and density make it a sub-optimal candidate for PET. 

However, cooling pure CsI to cryogenic temperatures, approximately 100~K, causes its light emission spectrum to shift toward the near ultraviolet range, moving from around 310 nm to 340 nm~\cite{amsler2002temperature}. This process also increases its light yield by approximately a factor of 20 to about 100~photons/keV, significantly improving its energy resolution~\cite{mikhailik2015luminescence, Lewis:2021cjv, Ding:2023pqe}. 

Unfortunately, its decay time also increases, approximately by a factor of 50. In a PET scanner, the coincidence time resolution (CTR) between pairs of crystals is determined by the material's emission time profile and amount of light detected. An approximate figure of merit is given by \cite{conti2009comparison}:

\begin{equation}
    \mathrm{CTR}\propto\sqrt{\frac{\tau}{N_{\mathrm{ph}}}},\label{eq:ctr}
\end{equation}
where $\tau$ is the decay time constant and $N_{\mathrm{ph}}$ is the number of detected photons.

Using eq.~\eqref{eq:ctr} to compare the CTR for CsI operating at room temperature and at 100 K we obtain:

\begin{align}\label{eq:ctrx}
    \mathrm{CTR}_{100~\mathrm{K}}  &\approx  \sqrt{\frac{800~\mathrm{ns}}{15~\mathrm{ns}}\cdot\frac{5~ \mathrm{ph./keV}}{100~\mathrm{ph./keV}}}\cdot  
    \mathrm{CTR}_{300~\mathrm{K}}  \\&\approx 1.6 \cdot \mathrm{CTR}_{300~\mathrm{K}}.\nonumber
\end{align}
Thus, given that the time resolution at room temperature is at the nanosecond level~\cite{kubota1988new}, timing should still be possible also at 100 K, in spite of the long decay time.  

We demonstrate experimentally that such high light yield translates into an energy resolution and a coincidence time resolution that satisfy the requirements of a PET scanner, pointing towards the potential of cryogenic CsI for this application. 

This paper is organized as follows. Section~\ref{sec:setup} describes the experimental setup, while section~\ref{sec:results} summarizes the results obtained in terms of light yield, energy resolution and coincidence time resolution. Conclusions and future prospects are summarized in section~\ref{sec:conclusions}.

\begin{figure*}[ht]
\centering
\subfloat[]{\includegraphics[width=0.55\linewidth]{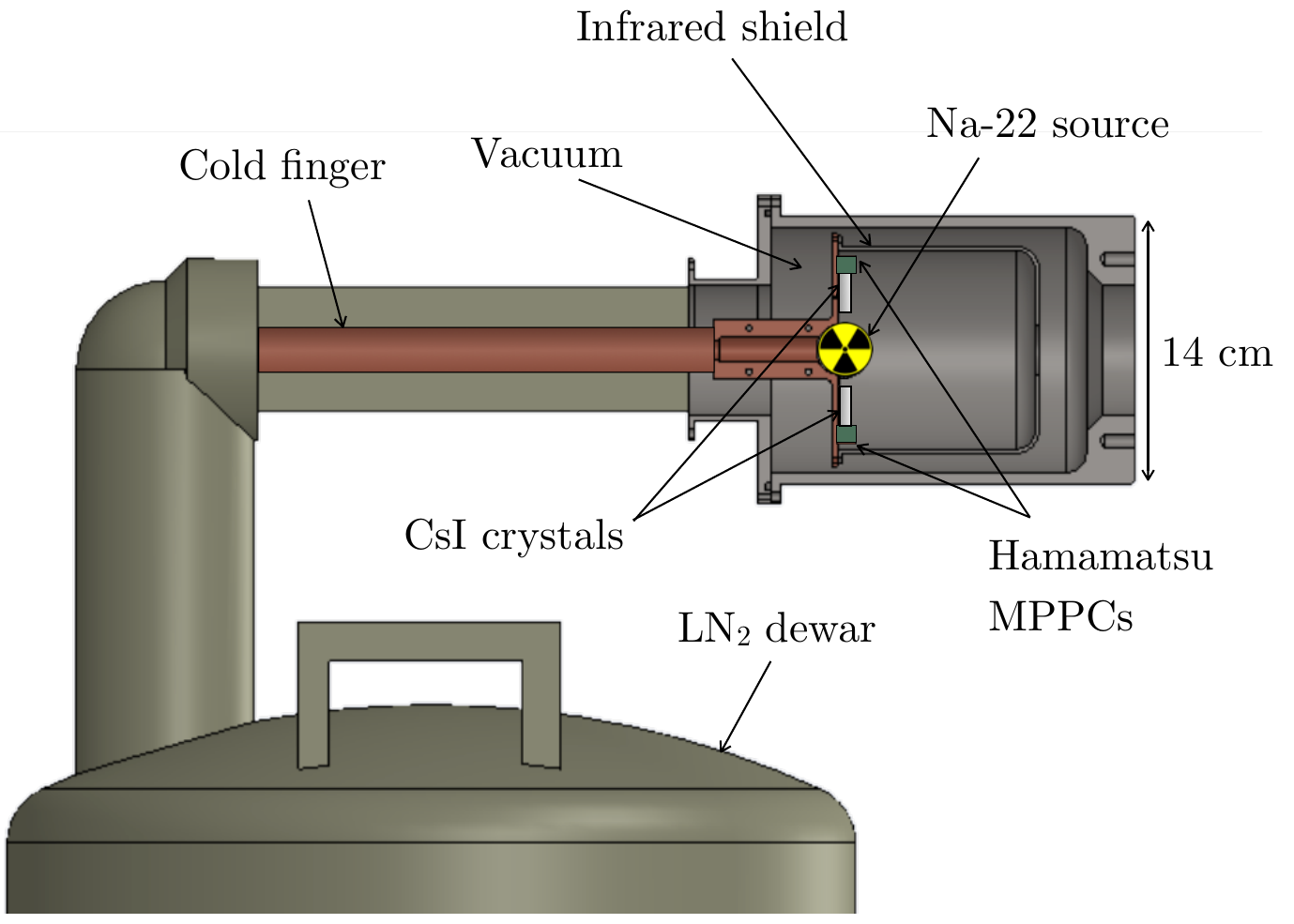}}
\hfil
\subfloat[]{\includegraphics[width=0.45\linewidth]{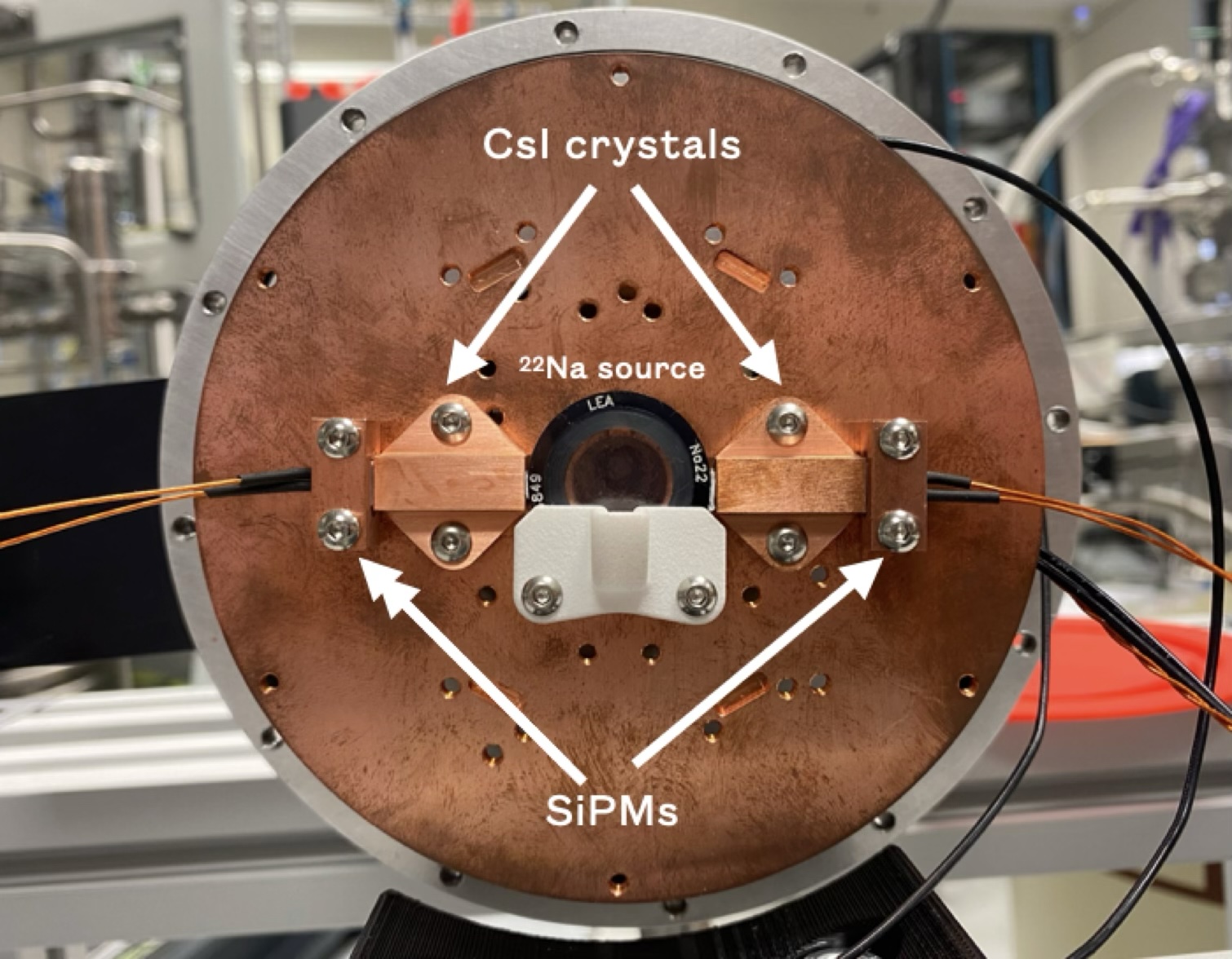}}
\caption{The crystals and photosensors are placed on a copper plate mounted inside a cryostat. The copper plate is connected to a cold finger that ca be immersed in liquid nitrogen. An infrared shield is placed on the copper plate to minimize radiative losses. (a) Schematic of the experimental apparatus. (b) Picture of the copper plate with photosensors and CsI crystals mounted on top.}\label{fig:setup}
\end{figure*}

\section{Experimental setup}\label{sec:setup}

We set up a table-top experiment, illustrated in figure~\ref{fig:setup}, employing a pair of pure CsI crystals of dimensions $3\times3\times20$~mm$^3$ provided by AMCRYS and polished on all sides. Their length corresponds to approximately 1.1~$X_{0}$. The crystals were wrapped with 3 layers of 80~\si{\micro\meter}-thick polytetrafluoroethylene (PTFE) tape and mounted on a copper plate which can be put in contact with a liquid nitrogen dewar through a cold finger. A Na-22 radioactive source with an activity of 40~kBq was placed between the two crystals. The light was detected by a pair of $3\times3$~mm$^2$ Hamamatsu S13360-3025CS MPPCs~\cite{Hamamatsu}. An aluminum infrared shield was placed above the copper plate to minimize radiative losses. The signals were acquired by a Tektronix MSO44 oscilloscope triggering the coincidence of the photosensors with a sampling rate of 3.125~Gs/s and a bandwidth of 200~MHz. Full waveforms were saved to disk with a trigger rate of approximately 5~Hz. 

A PT100 temperature sensor and two heating strips were placed behind the copper plate and connected to a CN32PT-224-C24 Omega PID controller, which allowed to keep the setup at the desired temperature. The setup was placed in a stainless steel chamber connected to a ConFlat 6-way cross. One flange featured 4 BNC connectors (2 per MPPC) and another one a 25-pin electronic feedthrough, used for the connection of the heating strips and the PT100 sensor. The chamber was then evacuated with an Agilent TwisTorr 305 fs turbopump to reach a vacuum level of approximately $10^{-4}$~mbar.

\section{Experimental measurements}\label{sec:results}
\subsection{Light yield and energy resolution}
The charge spectra of the MPPCs signals were obtained by integrating the full waveforms acquired by the oscilloscope. In order to convert the integral to the equivalent number of photoelectrons, the charge spectra were normalized by the single photoelectron charge peak, obtained by amplifying the dark pulses of the MPPC with a custom amplifier with a voltage gain of 600. 

\begin{figure}[htbp]
\centering
\includegraphics[width=0.7\linewidth]{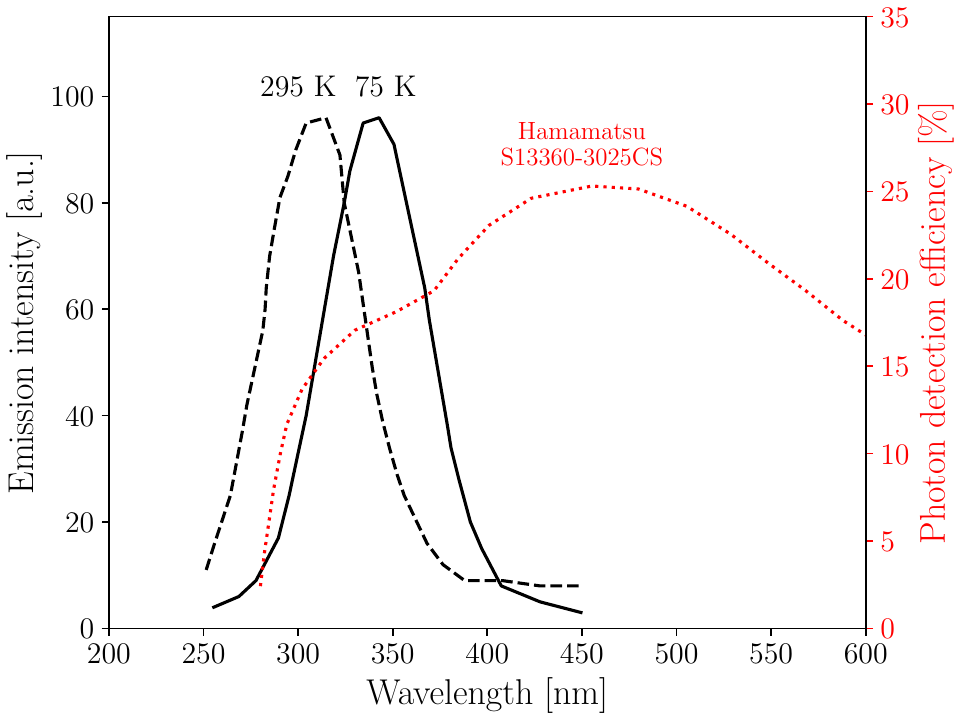}
\caption{CsI scintillation spectrum at room and cryogenic temperatures~\cite{amsler2002temperature} (in black, left axis), compared with the photon detection efficiency of the Hamamatsu S13360-3025CS MPPC~\cite{Hamamatsu} (in red, right axis).}\label{fig:csi_mppc}
\end{figure}

The intrinsic light yield of the crystal is obtained with the following formula: 
\begin{equation}
    \mathrm{LY}(\mathrm{photons/MeV}) = \frac{N_{\mathrm{pe}}/\mathrm{MeV}}{\mathrm{PDE}\times\epsilon_{\mathrm{det}}},
\end{equation}
where $N_{\mathrm{pe}}$ is the number of photoelectrons, $\mathrm{PDE}$ is the photon detection efficiency of the MPPC, and $\epsilon_{\mathrm{det}}$ is the detection efficiency of the setup. The weighted average PDE for the room temperature CsI emission spectrum is 12\%, compared with the 17\% for the CsI cryogenic spectrum (see figure~\ref{fig:csi_mppc}). However, conflicting measurements exist regarding the PDE of Hamamatsu MPPCs at cryogenic temperatures~\cite{Iwai:2019scy, Alvarez-Garrote:2024byb}, and no measurements is available for our specific model. Thus, a dedicated study is needed to improve the precision of this parameter for the estimation of the absolute light yield. 

The detection efficiency $\epsilon_{\mathrm{det}}$ has been estimated with a Geant4~\cite{GEANT4:2002zbu} Monte Carlo simulation using the \texttt{LUT} optical model~\cite{5485130} at 58\%. This relatively low efficiency is caused by the narrow form factor of the crystals (almost 7 times longer than wide), since the scintillation photons are reflected inside the crystals several times before reaching the photosensor. Thus, even if the PTFE reflectivity is approximately 95\%~\cite{janecek2012reflectivity}, the probability for a scintillation photon of being absorbed before reaching the MPPC compounds at every reflection, becoming non-negligible. Different optical models have also been explored (\texttt{UNIFIED}~\cite{levin1996more}, \texttt{DAVIS}~\cite{roncali2013simulation}) obtaining variations in the value of $\epsilon_{\mathrm{det}}$ smaller than 10\%.

A room temperature measurement was performed first, obtaining a light yield of $(4500\pm500)$~photons/MeV and $(4300\pm500)$~photons/MeV for the two crystals, respectively. The energy resolution was measured to be approximately 35\% FWHM at 511~keV for both. These values were obtained by fitting the 511~keV photopeak with a Gaussian distribution. 

After the dewar was filled with liquid nitrogen, the cold finger was inserted, and the system reached an equilibrium temperature of 104 K after approximately 5 hours.

The voltage breakdown of MPPCs is proportional to the temperature with a slope of approximately 54~mV/K~\cite{Hamamatsu}. At $T=104$~K, we adjusted the bias voltage by reducing it by 10.3~V to compensate for this effect. We then repeated the single photon measurement. The integration window was increased to 5.5~\si{\micro\second}, due to the longer decay time of CsI at low temperatures (see section~\ref{sec:time}). The MPPC dark count rate is halved every 5.3~K~\cite{NepomukOtte:2016ktf}, becoming negligible for our integration time at $T=104$~K. Under these low-temperature conditions, the number of photoelectrons $N_{\mathrm{pe}}$ increased by over an order of magnitude. This corresponds to light yields of $(91,000\pm2,000)$~photons/MeV and $(87,000\pm2,000)$~photons/MeV for the respective crystals.

\begin{figure}[htbp]
        \centering
        \includegraphics[width=0.7\linewidth]{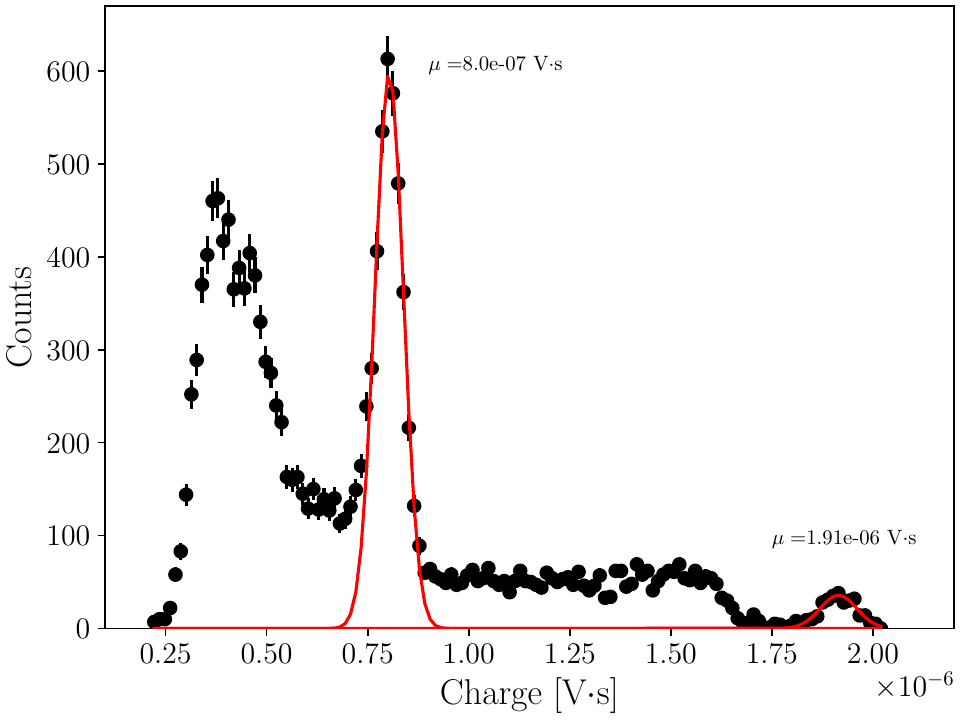}
        \caption{The charge spectrum was obtained by triggering a single channel and integrating the acquired waveforms. The red line corresponds to two Gaussian fits, one for each peak. Linearity of the response was verified by comparing the charge corresponding to the 511~keV peak with that of the 1275~keV peak.}
        \label{fig:linearity}
\end{figure}

Such a high light yield might affect the linearity of the response in a SiPM, as the probability for a photon to hit the same microcell during its recovery time is non-negligible. However, in the case of cryogenic CsI, the time emission constant is approximately 800~ns (see section~\ref{sec:time}), which spreads the photon emission over a relatively wide time window. Additionally, we are using SiPMs with 14,400 microcells (cell size is 25~\si{\micro\meter}). A conservative assumption on the cell recovery time is 100~ns~($2\times$ larger than the value obtained in ref.~\cite{grodzicka2015new}). Thus, with a light yield of 100,000~photons/MeV and a detector efficiency of 58\%, the photons that reach the sensor in the first 100~ns for an annihilation event are:
    \begin{equation}
        N_{t<100~\mathrm{ns}} = \frac{\int_0^{100~\mathrm{ns}} e^{-t/800~\mathrm{ns}} dt}{\int_0^{\infty} e^{-t/800~\mathrm{ns}}dt} \times 100,000 \times 0.511 \times 0.58 \approx 3500~\mathrm{photons.} 
    \end{equation}
With a PDE of 17\%, this corresponds to a number of fired cells of approximately 600 and an occupancy of 4.2\%, which is far from the saturation regime.
In addition, the response of the photosensor was verified to be linear by triggering only one channel: the charge corresponding to the Na-22 1275~keV peak is only slightly less than what one would extract by scaling linearly the charge at 511~keV, as shown in figure~\ref{fig:linearity}.

The setup was gradually warmed with the PID controller. The light yield and the single photoelectron peak for the two MPPCs were measured at various temperatures during this process. Each run lasted approximately 30 minutes and the MPPC bias voltage was adjusted accordingly for each temperature. The heating strips were able to warm the setup up to 180~K. Then, the cold finger was removed from the liquid nitrogen dewar and the cryostat was allowed to reach room temperature.

\begin{figure}[htbp]
\centering
\subfloat[]{\includegraphics[width=0.485\linewidth]{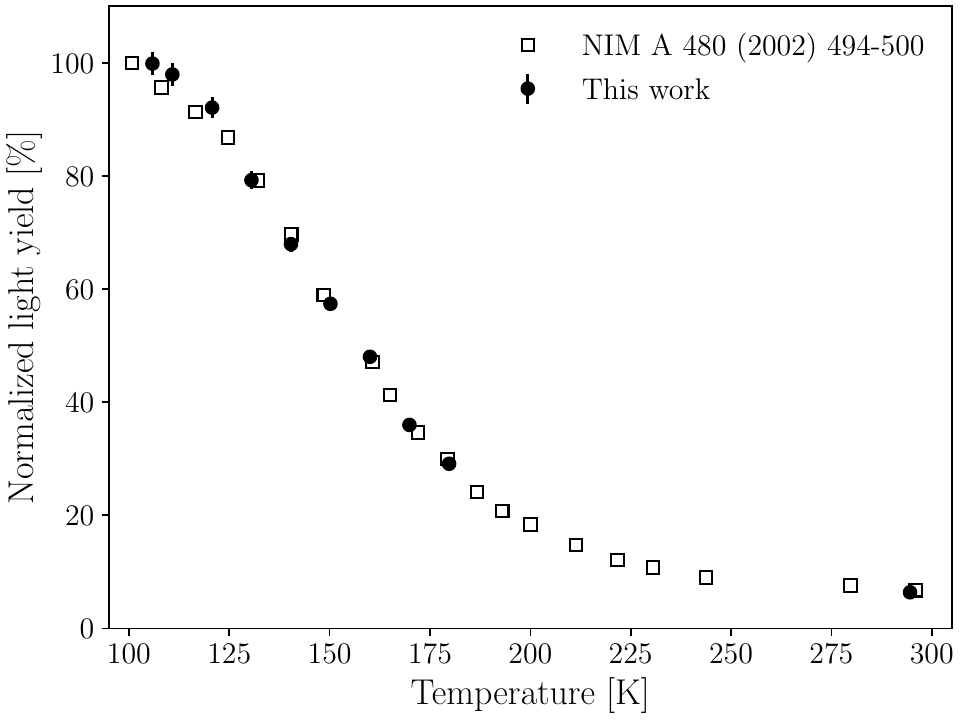}}
\hfil
\subfloat[]{\includegraphics[width=0.515\linewidth]{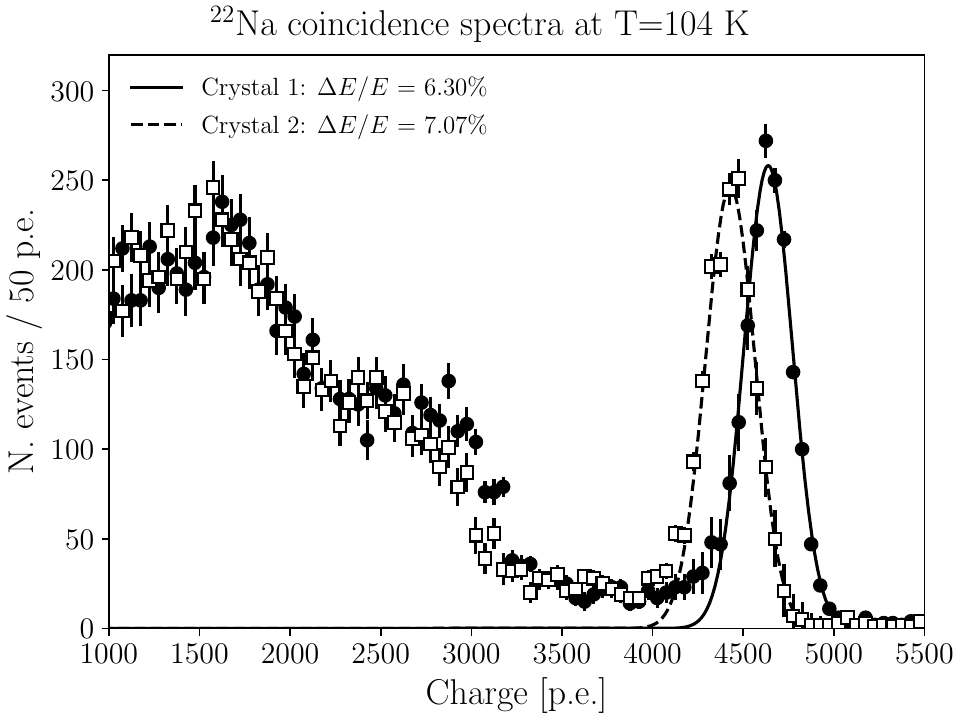}}

\caption{At cryogenic temperature, the CsI crystals show a 20-fold increase in the light yield, with a corresponding improvement in the energy resolution. (a) Light yield of pure CsI crystals as a function of temperature (black dots), compared with the existing literature~\cite{amsler2002temperature} (white squares). The heating strips were able to warm the setup up to 180~K, thus the absence of measurements between this value and the room temperature. (b) Na-22 charge coincidence spectra of the two CsI crystals operating at $T=104$~K. The 511~keV photopeak has been fitted with a Gaussian distribution (solid and dashed lines). }\label{fig:eres_cold}
\end{figure}

Figure~\ref{fig:eres_cold} (a) shows the non-linear increase in light yield as the temperature decreases, with the yield reaching a plateau at approximately 110 K, in good agreement with the existing literature~\cite{amsler2002temperature}. The approximate 20-fold increase between room temperature and $T=104$~K is mainly due to the larger amount of scintillation light being emitted. In addition, the light emission at cryogenic temperature is shifted towards larger wavelengths, where the quantum efficiency of the photosensor is higher (see figure~\ref{fig:csi_mppc}). This effect contributes to an additional 30\% increase in the amount of light detected.

This significant enhancement in light output translates into an improved energy resolution for the crystals, with values of $(7.07\pm0.05)\%$ and $(6.30\pm0.05)\%$, as illustrated in figure~\ref{fig:eres_cold} (b). 

For a scintillator coupled to a SiPM, the energy resolution can be expressed as~\cite{moszynski2005energy, NEXT:2024btf}:
\begin{equation}
    \Delta E/E = 2.355 \sqrt{\delta_{\mathrm{sc}}^2 + \frac{1}{N_{\mathrm{pe}}} + \frac{(\sigma_q/q)^2}{N_{\mathrm{pe}}} + \delta_{\mathrm{noise}}^2},\label{eq:resolution}
\end{equation}
where $\delta_{\mathrm{sc}}$ represents the intrinsic resolution of the crystal, $1/N_{\mathrm{pe}}$ is the photostatistic component, $\sigma_q/q$ is the relative variance of the photosensor, and $\delta_{\mathrm{noise}}$ is the dark noise component. In our setup, $\sigma_q/q$ was measured to be approximately 0.1, making its effect negligible. Similarly, $\delta_{\mathrm{noise}}$ can be considered negligible at cryogenic temperatures. 

By substituting the number of photoelectrons $N_{\mathrm{pe}}$ at 511~keV into this equation (4596 for crystal 1), we determined the intrinsic resolution of cryogenic CsI at this energy. Our result is $2.355\cdot\delta_{\mathrm{sc}}=5.3\%$, consistent with the value reported in the literature~\cite{moszynski2005energy}. 

It is important to notice that the SiPM we are using has a fairly low PDE (17\% for cryogenic CsI, see figure~\ref{fig:csi_mppc}), so the photostatistic term in eq.~\eqref{eq:resolution} is still significant. Thus, replacing it with a photosensor with a larger microcell size and a larger PDE (e.g., the Hamamatsu MPPC S13360-3075CS with 34\% PDE) should allow us to reach an energy resolution below 6\%. On the downside, adopting SiPMs with a larger microcell size (75~\si{\micro\meter} instead of 25~\si{\micro\meter}), and consequently a smaller number of microcells at parity of active area, might introduce a non-linear response which needs to be carefully evaluated~\cite{s24051671}.

\subsection{Emission time profile and coincidence time resolution}\label{sec:time}

\begin{figure}[ht]
\centering
\includegraphics[width=0.7\linewidth]{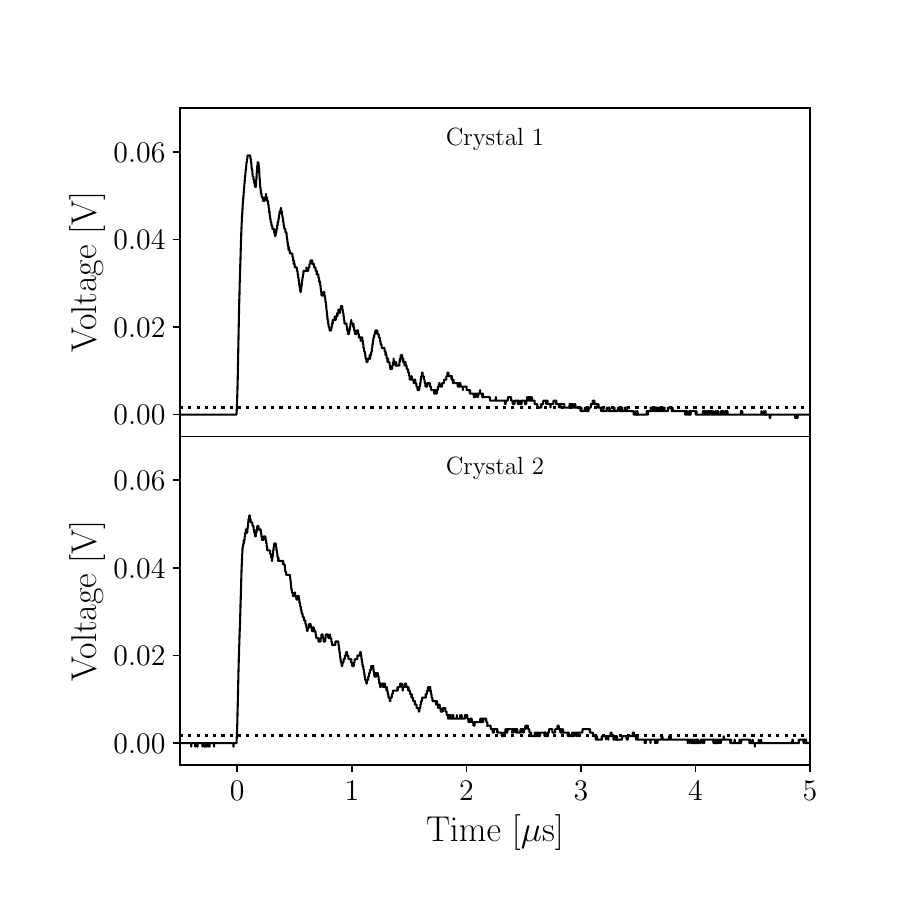}
\caption{An example of two waveforms acquired by triggering the coincidence of the two photosensors, with an equivalent energy near the 511~keV photopeak. The dotted lines correspond to the threshold used for the CTR measurement.}\label{fig:wf_example}
\end{figure}

The CTR for events near the 511~keV photopeak was measured in our setup using a baseline-corrected leading edge time pick-off method~\cite{du2017time}. The threshold was set at 5 times the electronic noise, defined as the standard deviation of the voltage values in a region with no signal and corresponding to approximately 3~mV. No improvement was observed with smaller threshold values and a worsening of the CTR was observed with a threshold lower than $3\sigma$. The events were selected by requiring a charge equivalent to the energy window $511~\mathrm{keV} - \mathrm{FWHM}/2 < E < 511~\mathrm{keV} + \mathrm{FWHM}/2 $. An example of a triggered event with an energy near the photopeak is shown in figure~\ref{fig:wf_example}.

\begin{figure*}[ht]
\centering
\subfloat[]{\includegraphics[width=0.56\linewidth]{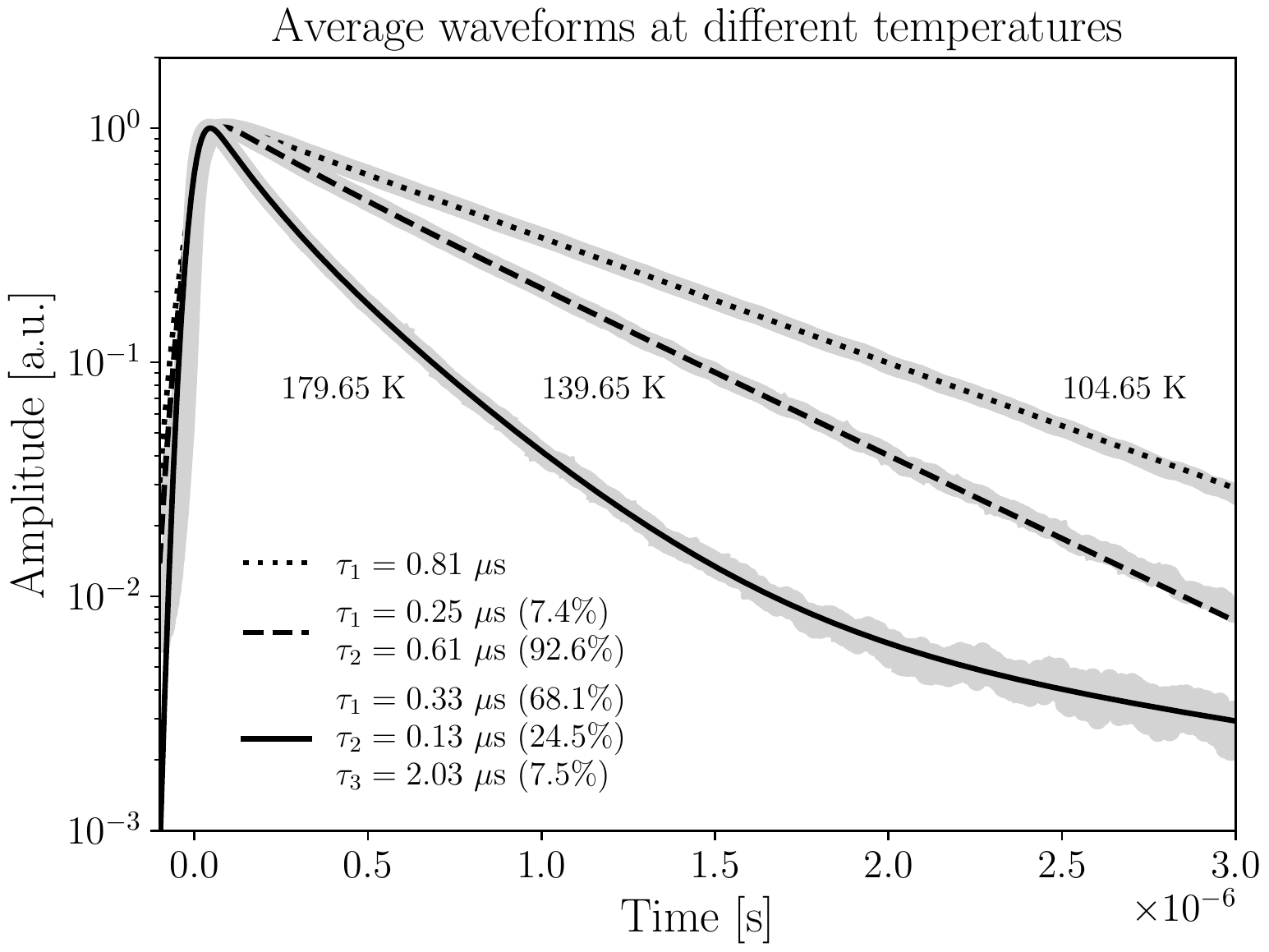}}
\hfil
\subfloat[]{\includegraphics[width=0.4\linewidth]{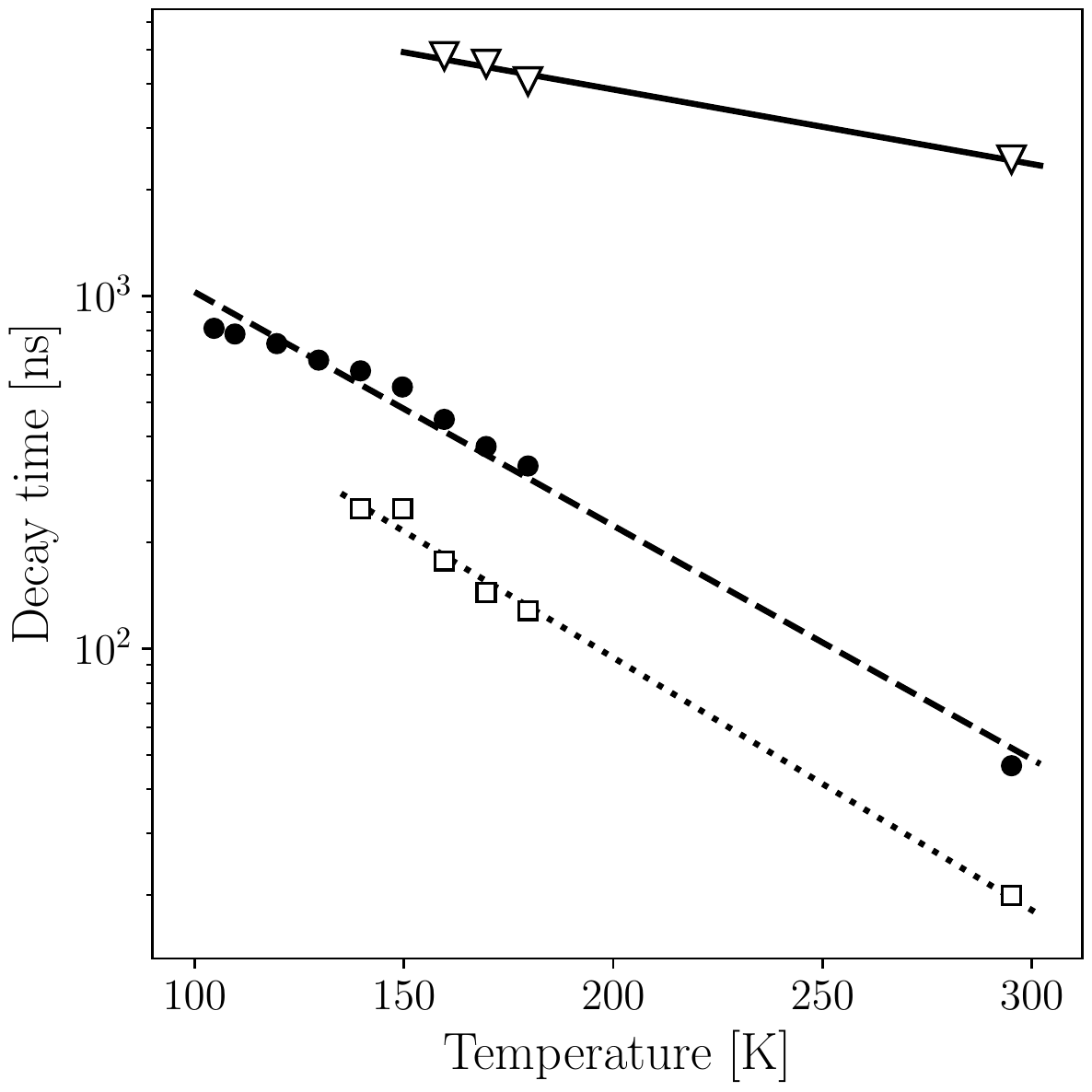}}
\caption{At cryogenic temperatures, pure CsI exhibits a time emission profile dominated by a single component of approximately 0.81~\si{\micro\second}. At higher temperatures two further components appear. (a) Average waveforms at three different temperatures fitted with one (dotted line), two (dashed line) or three (solid line) exponential functions. (b) Fast (white squares), medium (black circles) and slow (white triangles)  decay times as a function of temperature. The solid, dashed, and dotted lines represent linear fits in log scale.}\label{fig:time_profile}
\end{figure*}

At room temperature, a CTR of ($1.31\pm0.06$)~ns was achieved. The time emission profile is dominated by a fast component of approximately 15~ns, with two further components of approximately 50~ns and 2~\si{\micro\second} contributing to less than 10\% of the total light yield. By lowering the temperature, the values of the three components increase exponentially. Below 140~K, the fastest and the slowest component disappear, leaving a single component that reaches approximately 800~ns at $T=104$~K (see figure~\ref{fig:time_profile}).

\begin{figure*}[ht]
\centering
\subfloat[]{\includegraphics[width=0.49\linewidth]{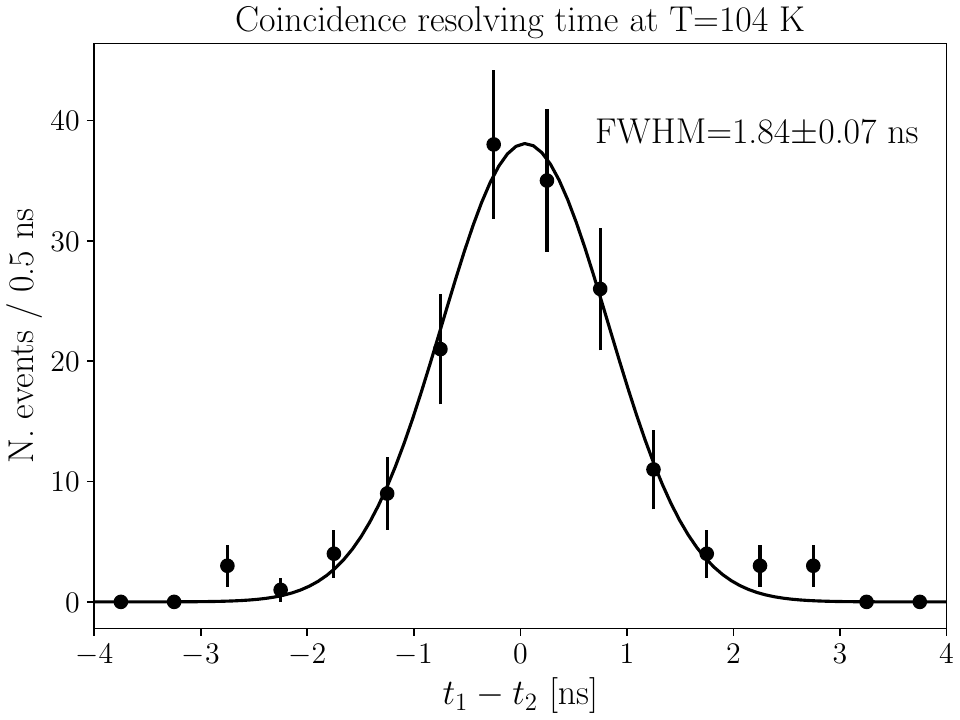}}
\hfil
\subfloat[]{\includegraphics[width=0.465\linewidth]{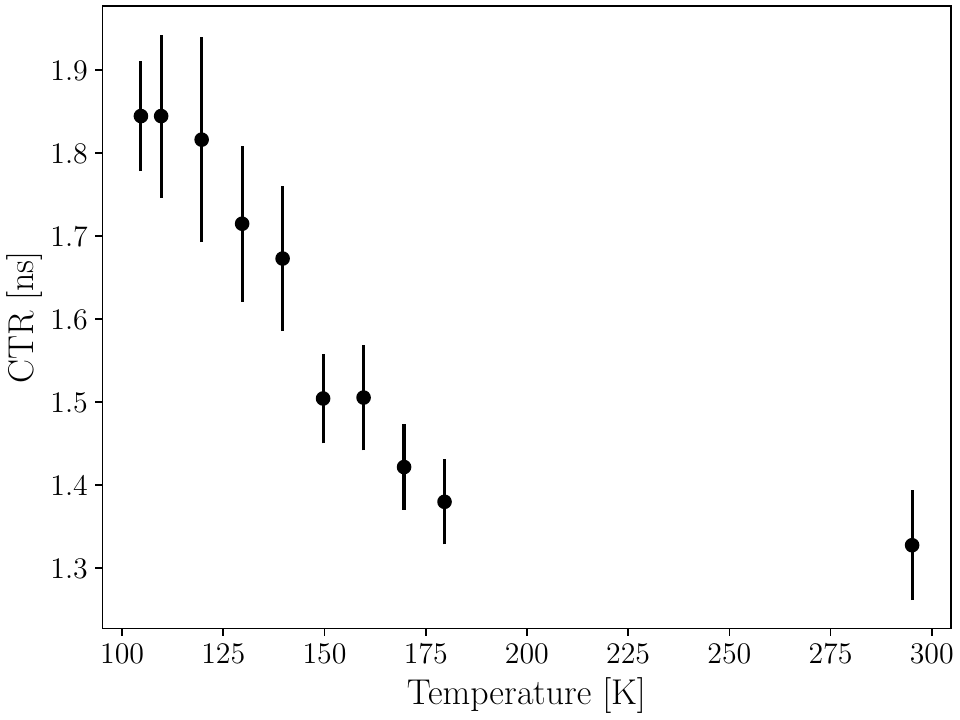}}
\caption{The coincidence time resolution increases with decreasing temperature, due to the larger decay time of CsI. The effect is partially compensated by the larger light yield. (a) Gaussian fit of the time difference at $T=104$~K. (b) CTR as a function of the temperature.}\label{fig:ctr}
\end{figure*}

At this temperature, the measured CTR was $(1.84\pm0.07)$~ns, which is qualitatively consistent with the value extrapolated from the room temperature measurement using eq.~\eqref{eq:ctr}. Thus, the effect of the increase in decay time is partially compensated by an increase in the number of detected photons, mitigating the impact on the CTR.
The CTR as a function of temperature and the time difference distribution at 104~K are shown in figure~\ref{fig:ctr}. Similarly to the case of the energy resolution, the value of CTR could be improved by adopting photosensors with a larger PDE, already available on the market. Following naively eq.~\eqref{eq:ctr}, a CTR of 1.3~ns could be achieved at cryogenic temperatures with a PDE approximately $2\times$ better. 

Further improvements might be possible by detecting Cherenkov photons, which are emitted promptly. In this context, CsI appears promising due to its low self-absorption and relatively high refractive index in the UV range (1.9 at 350~nm), where Cherenkov emission is more abundant. A preliminary simulation -- assuming a 75~\si{\micro\meter} cell MPPC (corresponding to a $2\times$ higher PDE) and a perfect time response from the electronics -- indicates that a CTR of approximately 400~ps could be achieved, potentially enabling time-of-flight (TOF) measurements.

\section{Summary and future prospects}\label{sec:conclusions}
We have characterized the response of a pair of pure CsI crystals read out by solid state photosensors (Hamamatsu MPPCs) in a setup cooled at cryogenic temperatures. At $T=104$~K, an energy resolution of 6.3\% and a coincidence time resolution of 1.84~ns have been achieved. The light yield and decay time have been measured as a function of temperature, with results in good agreement with the existing literature. 
For comparison, LYSO typically achieves energy resolutions around 7–10\%~\cite{ackermann2015time}, while BGO-based systems, such as those from GE~\cite{yamagishi2023performance}, can reach approximately 9\% at the system level.

The coincidence time resolution is sufficient for PET applications and might support TOF measurements with the detection of Cherenkov photons. Thus, this material might be a promising candidate for TBPET scanners, where the scintillator cost represents a large fraction of the total cost and can potentially offset the added complexity of a cryogenic system.

A full simulation of a PET scanner based on cryogenic CsI crystals will be developed, in order to evaluate the effect of an improved energy resolution and a lack of TOF on the image quality. 
Experimentally, the setup will be scaled up to two matrices of 64 SiPMs each, in order to test the electronics and the cryogenic system needed for a larger number of channels. Monolithic crystals will also be tested. This represents an interesting option, as it could enable the measurement of depth-of-interaction by using the spatial distribution of the collected light, thereby mitigating the drawback of the larger $X_0$ of the CsI.

Beyond its potential in PET, the properties of cryogenic CsI suggest its applicability to other medical imaging modalities. For instance, Single Photon Emission Computed Tomography (SPECT) could benefit from the enhanced spectral performance of cryogenic CsI. While the prolonged pulse duration increases detector dead time, potentially limiting count rates, this trade-off might be acceptable in SPECT systems where high energy discrimination is prioritized over timing. Further studies are needed to evaluate its practical implementation, including the optimization of readout electronics to mitigate the impact of the long decay time.

\section*{Acknowledgments}
We thank Paola Ferrario for the valuable feedback. SRS acknowledges the support of a fellowship from ``la Caixa Foundation'' (ID 100010434) with code LCF/BQ/PI22/11910019.

All authors declare that they have no known conflicts of interest in terms of competing financial interests or personal relationships that could have an influence or are relevant to the work reported in this paper.

\bibliographystyle{JHEP}
\bibliography{biblio.bib}

\vfill

\end{document}